\documentstyle[prb,twocolumn,aps]{revtex}

\begin{document}
\tolerance 50000
%%%%\preprint{
%%%%\begin{minipage}[t]{1.8in}
%%%%\hfill LPQTH-94/?
%%%%\end{minipage}
%%%%}

\draft

\title{Luttinger liquid behaviour and superconducting correlations
 in  t--J ladders}
\author{C.A.Hayward  and D.Poilblanc
}
\address{
Lab. de Physique Quantique, Universit\'e Paul Sabatier,
31062 Toulouse, France \\
}

\twocolumn[
\date{September 95}
\maketitle
\widetext
\vspace*{-1.0truecm}
\begin{abstract}
\begin{center}
\parbox{14cm}{
The low energy behaviour of the isotropic  t--J ladder system is investigated
using exact diagonalization techniques, specifically finding the
Drude weight, the charge velocity and the compressibility.
By applying the ideas of
Luttinger liquid theory, we  determine the correlation exponent
$K_\rho$ which  defines the  behaviour of the long range
correlations in the  system.
The boundary to phase separation is determined and a phase
diagram is presented.

At low electron density, a Tomonaga-Luttinger-like phase is stabilized
whilst at higher electron densities a gapped phase  with power
law pairing correlations is stabilized:
A large region of this gapped phase is found to exhibit dominant
superconducting
correlations.
}
\end{center}
\end{abstract}
\pacs{
\hspace{1.9cm}
}
]
\narrowtext

\section{Introduction}
Over the last few years, the behaviour of strongly correlated electrons
confined to coupled chains has received widespread attention; the
reasons for this are numerous. Firstly, the behaviour of electrons
in one-dimension, under t--J or Hubbard type interactions, is now
relatively well understood and described generally
by the term Luttinger Liquid (LL). The coupling of two
such Luttinger liquids as in the ladder geometry provides an interesting
first step towards the challenge of describing the behaviour
in two-dimensional systems. A second reason for interest in these
systems lies in the unusual nature of the ground state in the
undoped system, namely spin liquid behaviour with a finite gap
in the spin excitation spectrum \cite{heisladder};
this behaviour is in contrast to the
gapless behaviour of a single chain. The evolution of the spin-gapped
state on doping has obvious relevance to gapped superconducting behaviour.
In addition, compounds such as $(VO_2)P_2O_7$
 \cite{compounds1}
and $SrCu_2O_3$ \cite{compounds2}
are believed to be well described by a lattice of coupled chains.
Very recently, experiments on $La_{1-x}Sr_xCuO_{2.5}$ \cite{compounds3}
 have provided insight
into the doping of coupled chain systems.
Whilst there is considerable literature on many aspects of the
t--J ladder behaviour, a complete picture is still far from being
realised:
Our aim in this paper is to clarify some of the behaviour of the t--J
ladder system by drawing on some of the ideas used in describing
the strictly one-dimensional systems (i.e. LL theory). In conjunction
with results from several other techniques, we will then present
a speculative phase diagram.

The $t$-$J$ Hamiltonian on the $2\times L$ ladder is defined as,
\begin{eqnarray}
   {\cal H}=
   J^\prime \; \sum_{j}
   ({\bf S}_{j,1} \cdot {\bf S}_{j,2}
   - \textstyle{1\over4} n_{j,1} n_{j,2} ) \cr
   +J\;\sum_{\beta,j}
   ({\bf S}_{j,\beta} \cdot {\bf S}_{j+1,\beta}
   - \textstyle{1\over4} n_{j,\beta} n_{j+1,\beta} ) \cr
      - {t} \sum_{j,\beta ,s}
   P_G({c}^{\dagger}_{j,\beta;s}{c}_{j+1,\beta;s} + {\rm H.c.})P_G  \cr
      - {t^\prime} \sum_{j,s}
   P_G({c}^{\dagger}_{j,1;s}{c}_{j,2;s} + {\rm H.c.})P_G ,
\label{hamiltonian}
\end{eqnarray}
\noindent
where most notations are standard. $\beta$ (=1,2)
labels the two legs of the ladder (oriented along the
$x$-axis) while $j$ is a rung index ($j$=1,...,$L$).
We shall concentrate on the isotropic case where
the intra-ladder
 (along $x$) couplings   J and t  are equal to
the inter-ladder   (along $y$) couplings   $J^\prime$ and $t^\prime$.

At half filling the hamiltonian reduces to the Heisenberg model and the
behaviour is generally relatively well understood
\cite{heisladder}. A simple interpretation is given by considering the
strong coupling limit ($J=0$) in which the ground state consists of a
singlet on each rung with a spin gap ($\sim J^\prime$) which corresponds to
forming a triplet on one of the rungs. With the introduction of
intra-chain coupling $J$, the triplets can propogate and form a
coherent band thereby reducing the spin gap. In the isotropic case, the
spin gap remains ($\sim 0.5J$).

The evolution of this spin gapped state on doping is perhaps one of the
most interesting aspects of the ladder behaviour \cite{gapped}.
Recent work on this hole-doped  phase \cite{DP,chdpprl,ttr}
 has indicated a finite spin gap,
a single gapless charge mode, hole pairing and possible
dominant superconducting correlations. In this paper
we shall discuss some new independent results which
provide a more complete description of not only this gapped phase but
the whole region of parameter space.
A possible phase diagram  for the isotropic t--J ladder as a function of
$J/t$ and doping has been proposed recently \cite{DP} and we use these ideas in
our
analysis:  Away from half filling
the spin-gapped phase is stabilized
 up to
$J/t\sim 2.1$ where the system phase separates \cite{tsunetsugu}. As the system
is doped further, a phase with a single
gapless spin and a single  gapless charge mode is
found and,  as we shall explain, this behaviour is like that
of the one-dimensional Tomonaga-Luttinger-Liquid system. As in the gapped
phase,
phase separation occurs for large $J/t$. At very small electron densities,
an electron paired phase exists.
Note that although in the  t--J ladder
there are four possible zero-momentum gapless modes (two
spin and two charge), throughout the phase diagram
only one gapless charge mode
and either zero or one gapless spin modes are observed; this then
allows an almost identical treatment to the strictly one-dimensional
case.

In section \ref{LLB} we discuss briefly the Luttinger Liquid theory used to
describe strictly one-dimensional systems and also the possible application
of this theory to coupled chains; In section \ref{num} we present our
numerical calculations, and finally in section \ref{disc} we apply the
Luttinger Liquid theory to our results and present a phase diagram
for the system.

\section{Luttinger Liquid Behaviour: Tomonaga-Luttinger and Luther-Emery phases
}
\label{LLB}

In dealing with strictly one-dimensional interacting fermion systems,
one can in general make use of conformal field theory \cite{cft} and
bosonization \cite{boson} which allow a determination of the
decay exponents of the various correlation functions to be determined
from the low energy behaviour of the model.
The general idea  is that
one-dimensional interacting
fermion systems can be mapped onto the Fermi-gas model and the
corresponding `g-ology' weak coupling theory \cite{gology}.
This Fermi gas model scales to two different regimes,
namely the Tomonaga-Luttinger (TL)
fixed point and the Luther-Emery (LE) line which are
relevant for repulsive ($g_1>0$)
and attractive ($g_1<0$) backscattering matrix elements respectively;
as we shall explain, the important
difference between these two universality classes lies in the
spin degrees of freedom.
The low lying excitations of the Fermi gas model are collective
spin or charge density
oscillations, which propagate with different velocities, giving rise to
spin-charge separation and power law behaviour of the correlation functions.
In the TL phase, both  a gapless spin and a gapless charge mode are
exhibited; In contrast, whilst exhibiting a gapless charge mode,
the LE phase has a gap to spin excitations.

Conformal field theory relates the
properties of a finite system (such as the compressibility and the
Drude weight) with the correlation exponents;
One coefficient ($K_\rho$ ) determines the exponents of all  the
power law decays, (and
similarly the singularity of the momentum distribution function close to
$k_f$). Hence if a particular model scales to the TL (or LE) universality
class, one can infer the dominant correlations from the low energy behaviour
of the system which can be deduced from much smaller system sizes than would
be required to calculate the correlation lengths directly.

The relationships between the correlation exponent ($K_\rho$)
 and the low energy behaviour
of the model  are given below for a system of size N, and length L
( we have chosen to give general equations such that
N=L for a chain, N=2L for the ladder geometry).
Firstly, the ratio of the charge velocity $u_\rho$
 to the coefficient $K_\rho$
is proportional to the variation of the ground state energy $E_0$ with
particle density n, i.e. the inverse compressibility

\begin{eqnarray}
{\pi\over 2}{u_\rho\over K_\rho}={1\over n^2\kappa}=
{\partial^2(E_0/N)\over \partial n^2}
\label{lutt1}
\end{eqnarray}
\noindent
The coefficient $K_\rho$ is also related to the Drude weight
$\sigma_0$; this Drude weight is the weight of the zero
frequency (dc) peak in the conductivity $\sigma_\omega$
and may be obtained by considering the curvature of the ground state energy
level as a function of threaded flux, $\Phi$

\begin{eqnarray}
\sigma_0=2u_\rho K_\rho = {L^2\over 4\pi}{\partial^2(E_0/N)\over
\partial\Phi^2}
\label{lutt2}
\end{eqnarray}
\noindent
We can also determine the charge velocity by considering the dispersion
of the  energy spectrum
\begin{eqnarray}
u_\rho=(E_{1\rho}-E_0)/(2\pi /L)
\label{lutt3}
\end{eqnarray}
\noindent
where $E_{1\rho}$ is the lowest lying charge mode to the ground state ($E_0$)
with
neighbouring $k$ value.

These equations provide us with 3 independent conditions on $K_\rho$ and
$u_\rho$ which  can be  used  to check the
consistency of  the Luttinger liquid relations.
Also a calculation of the parameter $K_\rho$ is relatively straightforward
and this parameter then determines the exponent coefficients of all the
correlation functions.

As explained previously, the essential difference between the TL and LE fixed
points lies in the spin degrees of freedom: the LE region gapped whilst the
TL gapless. The correlation exponents of the two different cases are summarised
in  table I
(taken from reference \cite{marco1}).
We have omitted the
logarithmic corrections and these are detailed in reference \cite{log}.
We emphasise that for LL systems, the correlation functions show either
power law or exponential decay, with interaction-dependent powers determined by
one coefficient $K_\rho$.
Also we see that for $K_\rho <1$ (spin or charge) density waves at
$2k_f$ are enhanced and diverge, whereas for $K_\rho >1$ pairing
fluctuations dominate.

Whilst in strictly one-dimensional systems there is a single gapless
charge mode,
the theory could equally well be applied in a case where more than one
gapless mode existed if the excitations were decoupled in the low
energy regime.
In such a case, each degree of freedom would have an
associated operator algebra, i.e. an associated $K_\rho$.

Specifically considering
the problem of coupled chains, we note that at present little
is known  and there is much interest into how
to connect the quasi-one-dimensional
results to the strictly one-dimensional case. A recent calculation by Schulz
\cite{HJS}  has
considered the coupling of two Luttinger liquids by a small interchain hopping
using a bosonization technique. Interestingly, in the presence of both forward
and backward scattering terms,
 the calculation predicts a gap in all the magnetic
excitations and a gapless charge mode
(as observed for the hole-doped region of the t--J ladder).
This gapped phase  however
exhibits  somewhat different correlations  to
the strictly one-dimensional  LE phase
described above.
Firstly the $CDW_\pi$ and $SDW_\pi$ correlations decay exponentially along the
chains (we use the notation of Schulz where $0$ or $\pi$ indicate
the oscillations are in or out of phase between the two chains respectively,
i.e. `bonding' or `antibonding').
A divergent density-density  response, decaying as
$\sim \cos{\left (2(k_f^0+k_f^{\pi})r\right )} r^{-K_\rho}$,
exists
in analogy with the $4k_f$ oscillations of a single chain ($k_f^0$ and
$k_f^\pi$ refer to the Fermi points of the bonding and antibonding
quasi particle branches respectively);
we note that the coefficient $K_\rho$ differs by
a factor of two to that used by Schulz
since we have chosen to define $\sigma_0$ and  $\kappa$ as the Drude
weight and compressibility {\it per site} rather than per rung.
The superconducting correlations (cross chain pairing) decay as
$r^{-1/K_\rho}$ and exhibit a `d'-like character.
 Hence for this gapped phase,  we would expect
dominant superconducting
correlations for $K_\rho >1$.
In agreement with these findings, different approaches by
Troyer {\it et al} \cite{ttr}, Nagaosa \cite{nagaosa} and
and  Balents and Fisher \cite{fisher}
predict similar behaviour in the spin-gapped region: i.e. `Luther-Emery-like'
in the sense that there exist two order parameters (analogous to
the on-site pairing and $2k_f$ CDW in the LE class) whose exponents
obey a reciprocal relation: these correspond to the pair field correlations
and to a four fermion operator
$\left <n_B(r) n_B(0)\right >$, where $n_B$ is the density of `bosonic' hole
pairs
bound on a rung
(analogous to the $4k_f$ oscillations
of a single chain (i.e.$2k_f^0$+$2k_f^\pi$)).
In table II we summarize the correlation exponents predicted
for this spin gapped phase in the ladder
geometry.

In the following section we give details of calculations using these ideas
to characterize the behaviour of the t--J ladder, finding the
parameter $K_\rho$ and hence the dominant correlation functions.
We consider various electron densities and various ratios of J/t
to build up a speculative phase diagram.

\section{Numerical calculations}
\label{num}

Since we require information concerning
the low energy properties of the model, the dominant technique we
have employed is that of exact diagonalization of finite systems, specifically
$2\times 5$ and $2\times 10$ double chain rings.
Exact diagonalization techniques are particularly well
adapted to the investigation
of low energy modes since implementation of various  quantum numbers is
straightforward; the various excitation modes can be obtained by calculating
the
ground state energy in each symmetry sector.
The low energy modes of the system  are characterized firstly by their spin:
singlet and triplet excitations correspond to charge and spin modes
respectively.
It is also useful to consider the parity of the states under a reflection
in the symmetry axis of the ladder along the direction of the chains:
Even ($R_x=1$) or odd ($R_x=-1$) excitations corresponding to bonding (B) or
anti-bonding (A) modes respectively ($0$ and $\pi$ as used by Schulz
\cite{HJS}). Finally
the dispersion relation of each mode is determined by the momentum
$k_x$=$2\pi n$/$L$.

In order to ensure that the antiferromagnetic correlations are not frustrated
when
one goes around each chain, we have  chosen
the electron number to be  always a multiple
of four; our results then concern electron densities 0.4 and 0.8 for
both system sizes and in addition 0.2 and 0.6 for the
larger system size. The absolute ground state
is given by the boundary conditions that form a closed shell in the
non-interacting Fermi sea
(obtained by turning off the interaction, $J$)
 and these are used
in the calculations of $u_\rho$ and $\kappa$, specifically
anti-periodic boundary conditions for
$n <0.5$ and
periodic boundary conditions for
$n >0.5$.

\subsection{Drude Weight and Anomalous Flux Quantization}

The first calculation we present concerns  the Drude weight, defined by
equation \ref{lutt2}.
The numerical technique involves threading the double chain ring with a flux
$\Phi$ and studying the functional form of the ground state energy with respect
to
the threaded flux, namely $E_0(\Phi )$. In general $E_0(\Phi )$ consists
of a series of parabola, corresponding to the curves of the individual
many body states $E_n(\Phi )$: This envelope exhibits  a periodicity of
one, where we have chosen to measure the flux in units of the
flux quantum $\Phi_0 =hc/e$.
Note that the function $E_0(\Phi )$ also gives a quantitative
value of the superfluid density $D_s$, which is in general different
from $\sigma_0$ \cite{swz}.
The Drude weight corresponds to
curvature of a single ground-state many-body energy level
whilst the superfluid density corresponds to the curvature
of the envelope of the individual many-body states as a function of flux.
However, since the flux ($\phi_c$) at which another many body energy level
crosses
the zero-flux ground state energy level  varies as
$\phi_c\sim (hc/e)L^{1-d}$ (where $d$ is the dimension)\cite{swz}, in
one-dimension
$\phi_c$ is independent of L; there are only a finite number of energy level
crossings in the the thermodynamic limit and $\sigma_0$ and $D_s$ are equal (up
to a factor of $2\pi$).

In addition to the Drude weight and the superfluid density, the function
$E_0(\Phi )$ also yields information regarding the phenomenon of anomalous flux
quantization; this has been explained in a previous publication \cite{chdpprl}
so we mention it only briefly here. Whilst in general the
ground state envelope $E_0(\Phi )$  exhibits a periodicity of one, Byers
and Yang \cite{byers_yang} have shown that in the thermodynamic limit $E_0(\Phi
)$ exhibits
local minima at quantized values of flux, the separation of which is $1/n$
where
$n$ is the sum of the charges in the basic group. Hence, for a paired
superconducting state we would expect minima in $E_0(\Phi )$ at
intervals of $1/2$. These minima are related to the existence of
supercurrents which are trapped in metastable states corresponding to the
flux minima and are thus unable to decay away \cite{Schrieffer}.
 It should be mentioned that
this anomalous flux quantization (AFQ) is an indication of pairing and is
not in itself sufficient to imply a superconducting state.

Numerically the application of a flux through the double chain ring is
achieved by modifying the kinetic term of the hamiltonian such that
\begin{eqnarray}
c^\dagger_{j,\beta ;s}c_{j+1,\beta ;s}\mapsto c^\dagger_{j,\beta
;s}c_{j+1,\beta ;s}
e^{i2\pi \Phi\over L}
\label{phase}
\end{eqnarray}
\noindent
where $\Phi$ is the flux through the ring measured in units of $\Phi_0$.
Hence the application of a flux is numerically equivalent to a change in the
boundary conditions of the problem; $\Phi =0$ representing periodic and
$\Phi =1/2$ representing anti-periodic boundary conditions.
In the thermodynamic limit, $\sigma_0$ must be independent of the phase
introduced at
the ring boundary \cite{fye} and therefore
we consider the whole of the envelope $E_0(\Phi )$ as a function of
flux (in general consisting of several parabola).

Choosing the parameters $n=0.8$ and $J/t=0.5$, we show in
 figure \ref{f1}a(b)  all the  possible spin and charge modes
of the $2\times 5$ ($2\times 10$) system, for all possible momenta,
as a function of applied flux. In the case of the larger system,
we show the full spectrum for $\Phi <0.25$ in order to simplify the diagram
(this work has been previously published \cite{chdpprl}).
For both system sizes the minimum energy is formed by charge (spin zero)
bonding modes; the excited modes with different quantum numbers
move further from the ground state as the system size is increased (a
result we have checked by finite size scaling) and hence
will not interfere with $E_0(\Phi )$. The existence of minima at
intervals of half a flux quantum (i.e. anomalous flux quantization)
clearly indicates the existence of pairing).

The envelope $L[E_0(\Phi )-E_0(\Phi =0)]$ has been extracted and is shown in
figure \ref{f2}a(b) along with equivalent plots from other regions of the phase
diagram: Figure \ref{f2}a shows the data for $J/t =1.0$ and electron densities
of
$0.4$ and $0.8$, whilst figure \ref{f2}b shows the data for an electron
density of 0.8 for ratios $J/t=0.5$ and 4.0.
Apart from the curve with $n=0.8$, $J/t=4.0$ (which appears to
scale to a flat function, consistent with a phase separated state),
all the data appears to
show only small  finite size effects.  Anomalous flux quantization is
observed for the larger electron density (for the lower values of $J/t$)
indicating pairing, and hence consistent with a superfluid state in this
region.

In order to determine the Drude weight, we simply calculate the average value
of
the curvature of  $L[E_0(\Phi )-E_0(\Phi =0)]$
over all $\Phi$; a quadratic curve was fitted to each portion.
In figure \ref{f3}a we plot the Drude weight as a function of $J/t$
for electron densities 0.2,0.4,0.6 and 0.8 for the $2\times 10$ system
and electron densities 0.4 and 0.8 for the $2\times 5$ system.
The curves are plotted up to a maximum in $J/t$ which is determined by
the value at which the system phase separates (see compressibility).
In figure \ref{f3}b we plot the Drude weight as a function of electron
density for various values of the ratio $J/t$.

There are several features of the resulting behaviour
we should mention:
Note firstly  that
finite size effects are relatively small with the $2\times 5$ results close to
those of the $2\times 10$ results.
The Drude weight increases as the electron density is increased from
zero, until it reaches one-quarter filling, then decreases with
increasing electron density.
Also as we would expect for a spin charge separated state,
the Drude weight is effectively independent of
$J$.

\subsection{Charge Velocity}

The second quantity we have  calculated is the charge velocity, defined by
equation
\ref{lutt3}. Considering the charge bonding modes (the lowest lying charge
modes),
the energy difference between the ground state energy level and the energy
level with
neighbouring momentum,
$\Delta k_x={2\pi\over L}$  was calculated. As an example, we show
in figure \ref{f4}a
the charge bonding modes for the case
$n=0.4$ $J/t=2.0$, indicating with solid lines the
specific energy levels whose energy difference gives the charge velocity.
We  note that the gap (and therefore $u_\rho$)  is approximately
constant as a function of flux.
For consistency with future data however,
we have calculated the charge velocity at the particular flux which gives
the absolute ground state, i.e. $\Phi =0.5$ for $n<0.5$ and
$\Phi =0$ for $n>0.5$;
The results are shown in figure \ref{f4}b  as a
function of $J/t$ for various system sizes and various electron
densities (the results for  $n=0.6$ are not included
since the numerics present some difficulty close to one-quarter filling).
Note again that finite size effects are relatively
small.

\subsection{Compressibility}

The next quantity we calculate is that of the compressibility, defined
by equation \ref{lutt1}. The finite size equaivalent is given by
\begin{eqnarray}
{1\over n^2\kappa}
\simeq{1\over 2L}\left [{ E_0(n+\Delta n)+E_0(n-\Delta n)-2E_0(n)
\over (\Delta n)^2}\right ]
\label{compress}
\end{eqnarray}
\noindent
where $E_0(n)$ is the ground state energy of the finite
 system of ladder length $L$
with
an electron density $n$.
As for the calculation of the charge velocity, the boundary
conditions have been chosen to give the absolute ground state.
$\Delta n$ represents the
 finite change in electron density, 0.2 and 0.4 for the $2\times 10$ and
$2\times 5$ systems respectively.
In
Fig.\ \ref{f5} we show the results of the calculation of 1/$n^2\kappa $
for electron densities corresponding to 0.2,0.4,0.6 and 0.8 for the
$2\times 10$ ladder and we also plot the result of the $2\times 5$ system for
an electron density of 0.4.

In addition to the determination of the specific values of the
inverse compressibility, the boundary to phase separation may also
be determined from these results.
It is well known that at sufficiently large values of $J/t$,
a system will undergo a separation into two phases; a hole-rich
and an electron rich phase. This effect arises principally to minimize the
number of broken antiferromagnetic bonds in the system.
Phase separation occurs when the compressibility diverges (i.e. a
`liquid' to a `solid' phase) and hence
the inverse compressibility vanishes.
Whilst there remain some small finite size effects, the general form
of the phase separation line is readily observed from figure \ref{f5}.
As electron density is increased,
the  value of J/t at which phase separation occurs
 decreases (although electron densities of 0.6 and 0.8 are both indicate phase
 separation close to J/t $\sim$ 2.1). These results agree well with those of
 Tsunetsugu et al\cite{tsunetsugu} who used a similar technique but varied both
system size and electron density simultaneously.
The phase separation curve can be extrapolated to all electron densities
and will be shown in the predicted phase diagram, figure \ref{f8}.

\section{Luttinger Liquid parameters for the  ladder}
\label{disc}

In order to explore the validity of the Luttinger liquid relations
in our problem, we consider the ratio
$\sigma_0/\pi n^2 \kappa u_\rho^2$ which  equals unity for
a Luttinger liquid. The results of the numerical calculation of this
quantity are shown in figure \ref{f6} for various electron densities.

At low electron densities (i.e. for the cases
$n=0.2$ and
$n=0.4$), previous work has suggested that a
TL phase is stabilized and the results of this ratio show good
agreement with the predicted value of unity; for the case of
$n=0.4$ an increase in system size shows the
ratio scaling towards unity.
For the hole-doped spin-gapped region
($n=0.8$), where the behaviour is less well understood,
 the system phase separates at a much lower value of $J/t$ and hence the
curve drops to zero at $J/t \sim 2.1$. Before this phase separation, the data
is not inconsistent with the `new' LE-like behaviour which may be
described by  equations \ref{lutt1}-\ref{lutt3}
(finite size effects are
largest for high electron density and low $J/t$).
Since the ratio is close to unity, it appears to confirm the
earlier justification for using this one-dimensional
theory, i.e. only a single  gapless charge mode is observed.

With the values of the compressibility and the Drude weight, we can obtain
an estimate of the coefficient $K_\rho$ using
$K_\rho =1/2\sqrt{\pi n^2\kappa\sigma_0}$.
 This behaviour of $K_\rho$ for the
different electron densities of the $2\times 10$ ladder is shown in
Fig.\ \ref{f7} and we have also plotted the results for a $2\times 5$ ladder
with electron density 0.4.
Note that as $J/t$ is increased, $K_\rho$ increases  (for all
electron densities)  becoming infinite at phase separation as
the compressibility diverges.
A similar calculation has been performed by Troyer {\it et al}
\cite{ttr,tsunetsugu} for a specific electron density of 0.857,
i.e. two holes on a $2\times 7$
ladder  and the results are consistent in this region.

Before analysing the behaviour further, specifically the importance of
$K_\rho$,
we present a speculative phase diagram of the isotropic t--J ladder
as a function of $J/t$ and electron density.
This phase diagram is shown in figure \ref{f8} and we discuss briefly
the various regions.

At larger values of $J/t$, the system phase separates, and
to estimate the value of $J/t$ at which this occurs,
we show  the  data points at which the inverse
compressibility vanishes in the $2\times 10$ system (see figure \ref{f5}).
On doping away from half filling, the spin gap region persists and
a phase exhibiting one gapless charge mode is stabilized. On further doping
a gapless phase is stabilized: the schematic  boundary between these
two phases is shown as a dot-dashed line.
As for both the one- and two- dimensional t--J cases \cite{marco},
a gas of electron pairs is formed at low electron densities  above a
critical value of the ratio J/t; other 2p-particle (p$>$1) boundstates could
also become stable at larger J/t in this region.
Again the boundary to this paired phase is dot-dashed and is schematic.

 From the data in figure \ref{f7} we have plotted contours of
constant $K_\rho$ in order to allow a determination
of the dominant correlation functions.
Whilst the parameter $K_\rho$ is  continuous for
a particular value of J/t as the electron density is
varied \cite{troyer}, at some region between $n=0.4$ and
$n=0.8$ a gap opens in the spin excitation spectrum and
the correlation functions
 change their form discontinuously, scaling to a
different fixed point as explained in section \ref{LLB}.
A `jump' in the exponent of the superconducting correlations occurs with
the SCd charge exponent changing from $1+1/K_\rho$ to $1/K_\rho$; the $2k_f$
CDW
correlations jump from power law behaviour (exponent $1+K_\rho$) to
exponential decay.
In addition, in  the gapped high density state we  would
expect  conjugate
`four fermion $4k_f$' CDW correlations  with exponent $K_\rho$.

For both the high density `gapped' phase and the low density TL phase
we expect different correlation functions to dominate either side
of the contour $K_\rho =1$; in both cases superconducting correlations dominate
for $K_\rho >1$.
However, in the gapped hole-doped phase, this region is much larger
and in contrast to other models such as the one-dimensional
t--J model \cite{marco}, does not just exist as a
precursor to phase separation.
A physical picture of the behaviour is
of
the holes pairing  up on the rungs, the
 spins in singlets, and the dominant correlation  functions are then
 associated with
the movement of the hole pairs.

\bigskip
We wish to thank H.J.Schulz for many useful discussions and comments;
we also  gratefully acknowledge many helpful conversations  with
M.Luchini, F.Mila, W.Hanke and D.J.Scalapino.
{\it Laboratoire de Physique Quantique, Toulouse} is
{\it Unit\'e de Recherche Associ\'e au CNRS No 505}.
CAH and
DP  acknowledge support from the EEC Human Capital and Mobility
program under Grants ERBCHBICT941392 and  CHRX-CT93-0332.
We also thank IDRIS (Orsay)
for allocation of CPU time on the C94 and C98 CRAY supercomputers.

%
%  Fig. 1
%
\begin{figure}
\caption{
Energy as a function of flux (in units of $\Phi_0=hc/e$) for a) the $2\times 5$
 and b) the $2\times 10$ ladders with
$J/t=0.5$ and $n=0.8$. We show all possible momenta for various
quantum numbers: For the charge modes, the solid lines correspond to bonding
and
the dotted lines to anti-bonding, while for the spin modes the dashed lines
correspond to bonding and the dot-dashed lines to anti-bonding. For the
larger system size, we give only the charge bonding and the lowest lying spin
antibonding mode in full to simplify the diagram.
\label{f1}
}
\end{figure}
%
%  Fig. 2
%
\begin{figure}
\caption{
$L[E_0(\Phi )-E_0(\Phi =0)]$ where $L$ is the length of the ladder and
$E_0(\Phi )$
is the ground state energy with an applied flux $\Phi$. The dashed lines
correspond
to $2\times 5$, the solid lines to $2\times 10$. a) shows the results for
$n= 0.4$ and
$n= 0.8$ both with $J/t=1.0$, while b) shows
$J/t=0.5$ and $J/t=4.0$ both with
$n= 0.8$.
\label{f2}
}
\end{figure}
%
%  Fig. 3
%
\begin{figure}
\caption{
The Drude weight $\sigma_0$ a) as a function of ratio $J/t$  for
various electron densities and system sizes, and b) as a function
of electron density for various $J/t$.
The smooth lines are
a guide to the eye.
\label{f3}
}
\end{figure}
%
%  Fig. 4
%
\begin{figure}
\caption{
(a) Energy of the charge bonding modes as a function of flux for the
$2\times 10$ system with
$n=0.4$,$J/t=2.0$. The solid lines indicate the ground state and the
excited states used to calculate the charge velocity.
(b)
The charge velocity $u_\rho$ as a function of ratio $J/t$ for various
electron densities and system sizes. The dotted lines are a guide
to the eye.
\label{f4}
}
\end{figure}
%
%  Fig. 5
%
\begin{figure}
\caption{
$1/n^2\kappa$
as a function of ratio $J/t$ for various
electron densities and system sizes. The dotted lines are a guide
to the eye.
\label{f5}
}
\end{figure}
%
%  Fig. 6
%
\begin{figure}
\caption{
The ratio $\pi n^2 \kappa u_\rho^2 /\sigma_0$
as a function of $J/t$ for various
electron densities and system sizes. The dotted lines are a guide
to the eye.
\label{f6}
}
\end{figure}
%
%  Fig. 7
%
\begin{figure}
\caption{
$K_\rho$
as a function of ratio $J/t$ for various
electron densities and system sizes. The dotted lines are a guide
to the eye.
\label{f7}
}
\end{figure}
%
%  Fig. 8
%
\begin{figure}
\caption{
Speculative phase diagram of the t--J ladder as a function
of $J/t$ and electron density $n$. The circles and crosses represent results
from the $2\times 10$ system and the dotted lines are guides to the eye.
The dot-dashed lines separating the different phases are estimated.
\label{f8}
}
\end{figure}

%
% Table 1
%
\begin{table}
\caption{
Correlation exponents for the Tomonaga Luttinger and Luther Emery cases.
SS and TS indicate singlet and triplet superconductivity (pairing)
correlations respectively.
}
\begin{tabular}{|c|c|c|}
Correlation &TL &LE\\ \hline
2$k_f$ SDW & $1+K_\rho$ & exponential \\
2$k_f$ CDW & $1+K_\rho$ & $K_\rho$ \\
SS & $1+1/K_\rho$ & $1/K_\rho$ \\
TS & $1+1/K_\rho$ & exponential \\
4$k_f$ CDW & $4K_\rho$ & $4K_\rho$ \\
\end{tabular}
\end{table}

%
% Table 2
%
\begin{table}
\caption{
Correlation exponents for the spin  gapped  phase in the ladder geometry.
The  $\pi$ indicates that the correlations along the two chains
are out of phase.
}
\begin{tabular}{|c|c|}
Correlation & gapped spin\\ \hline
2$k_f$ SDW$_\pi$ & exponential \\
2$k_f$ CDW$_\pi$ & exponential \\
SCd & 1/$K_\rho$  \\
4$k_f$ CDW & $K_\rho$ \\
\end{tabular}
\end{table}

\begin{references}
\bibitem{gapped} The effect of doping spin gapped systems and the possible
presence of superconducting correlations has been studied previously in
various one-dimensional models eg t-J-J$^\prime$ hamiltonian \cite{marco1}
and the
dimerized t-J model \cite{imada}.
\end{references}
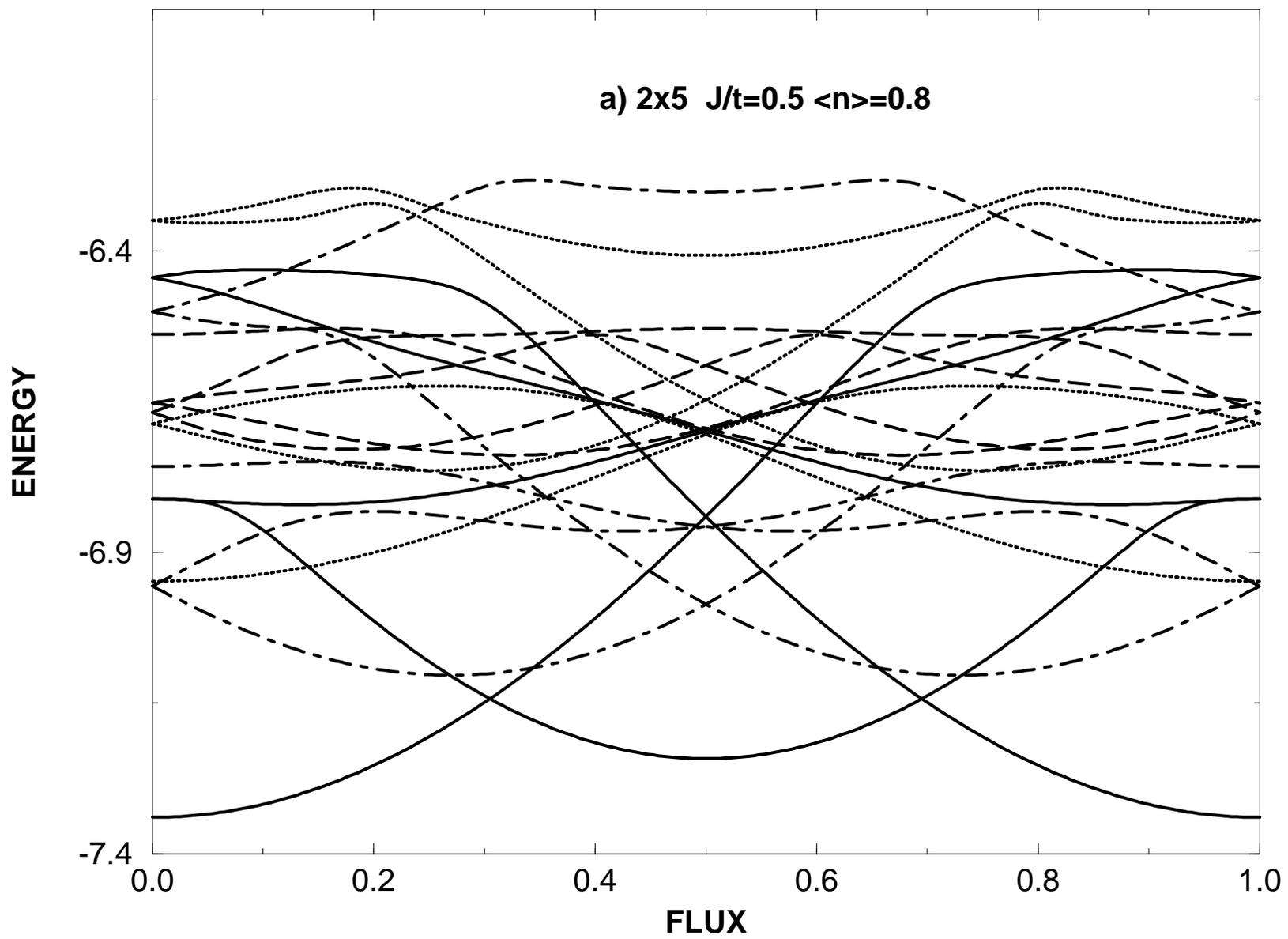
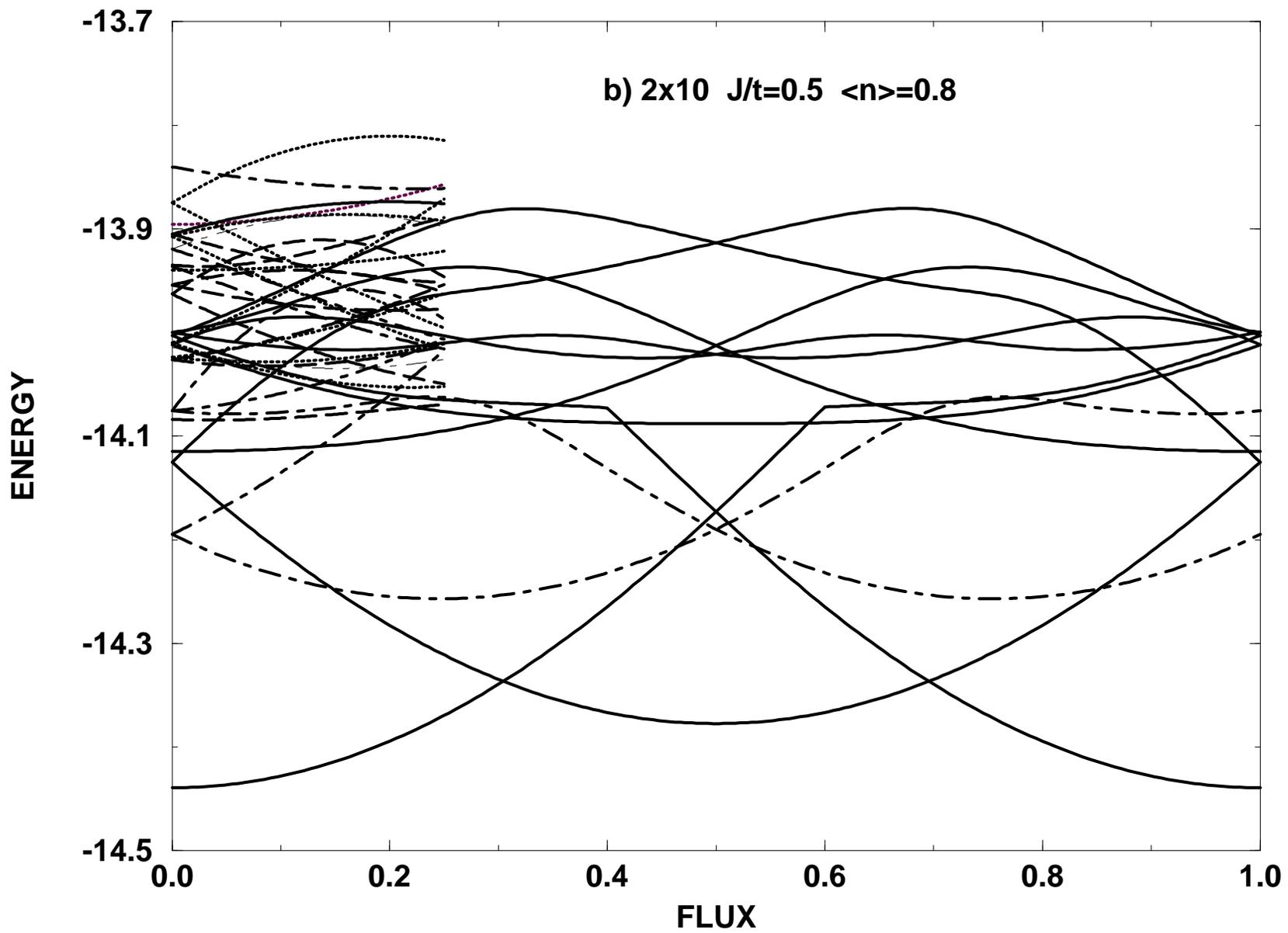
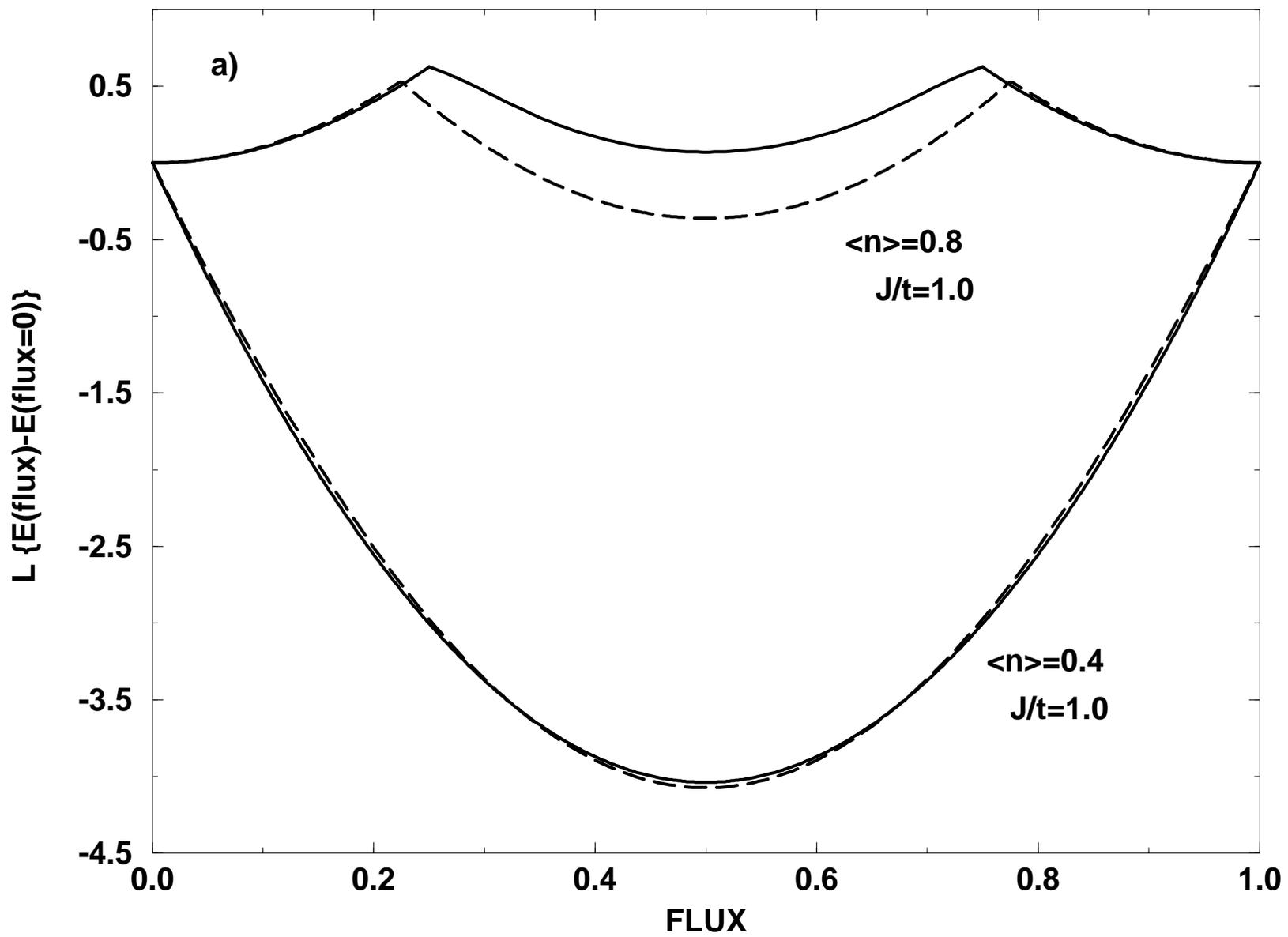
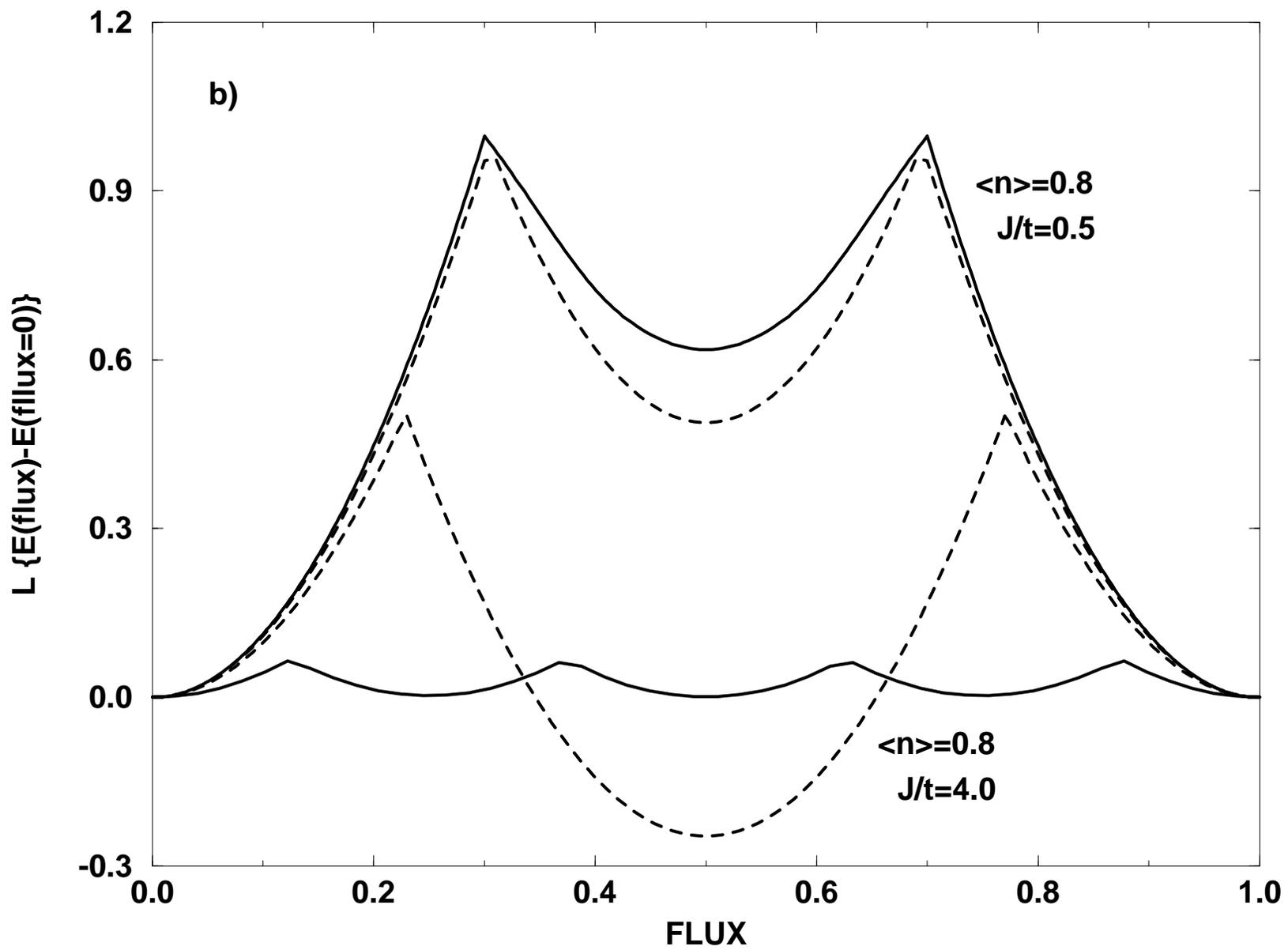
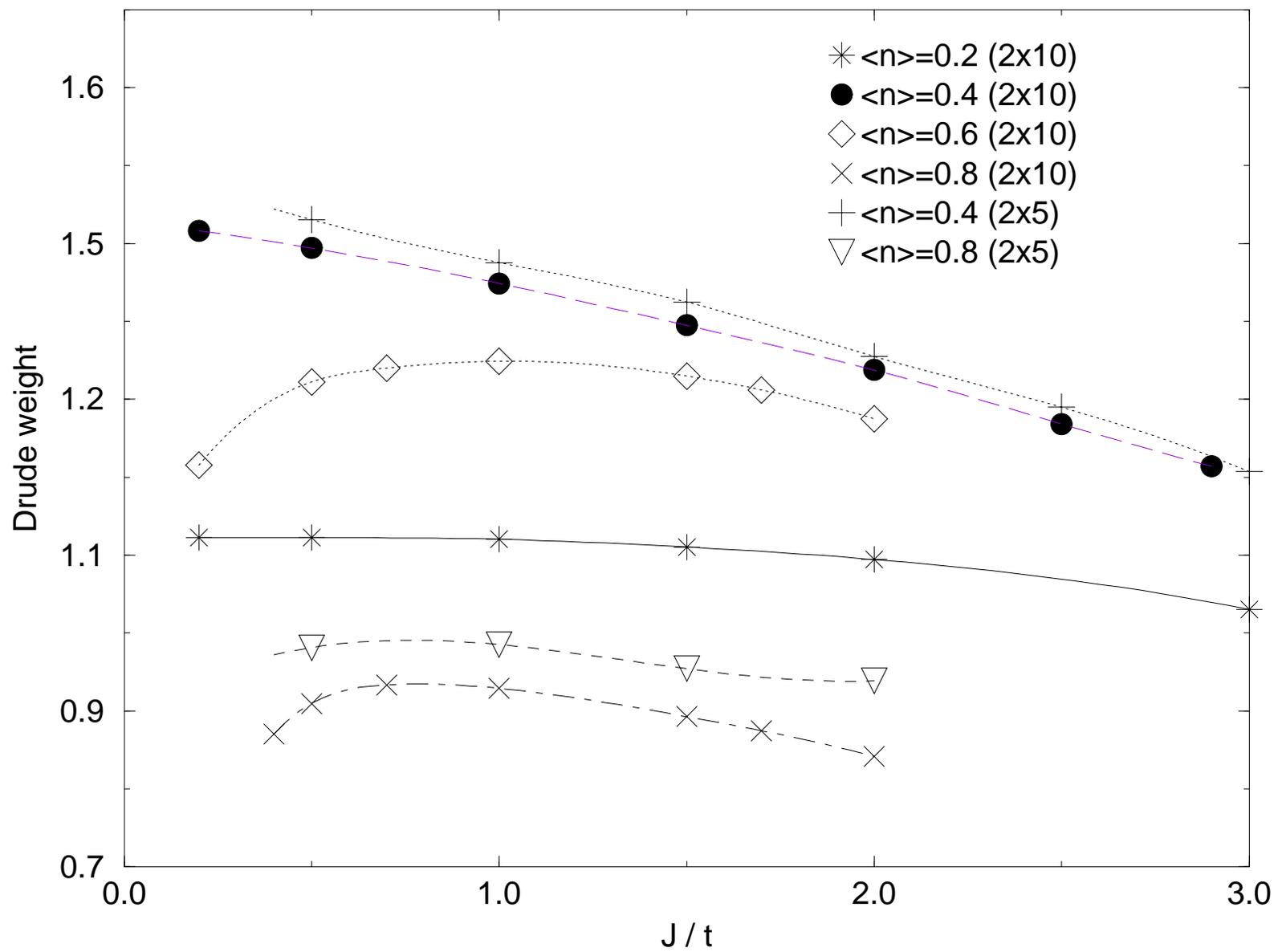
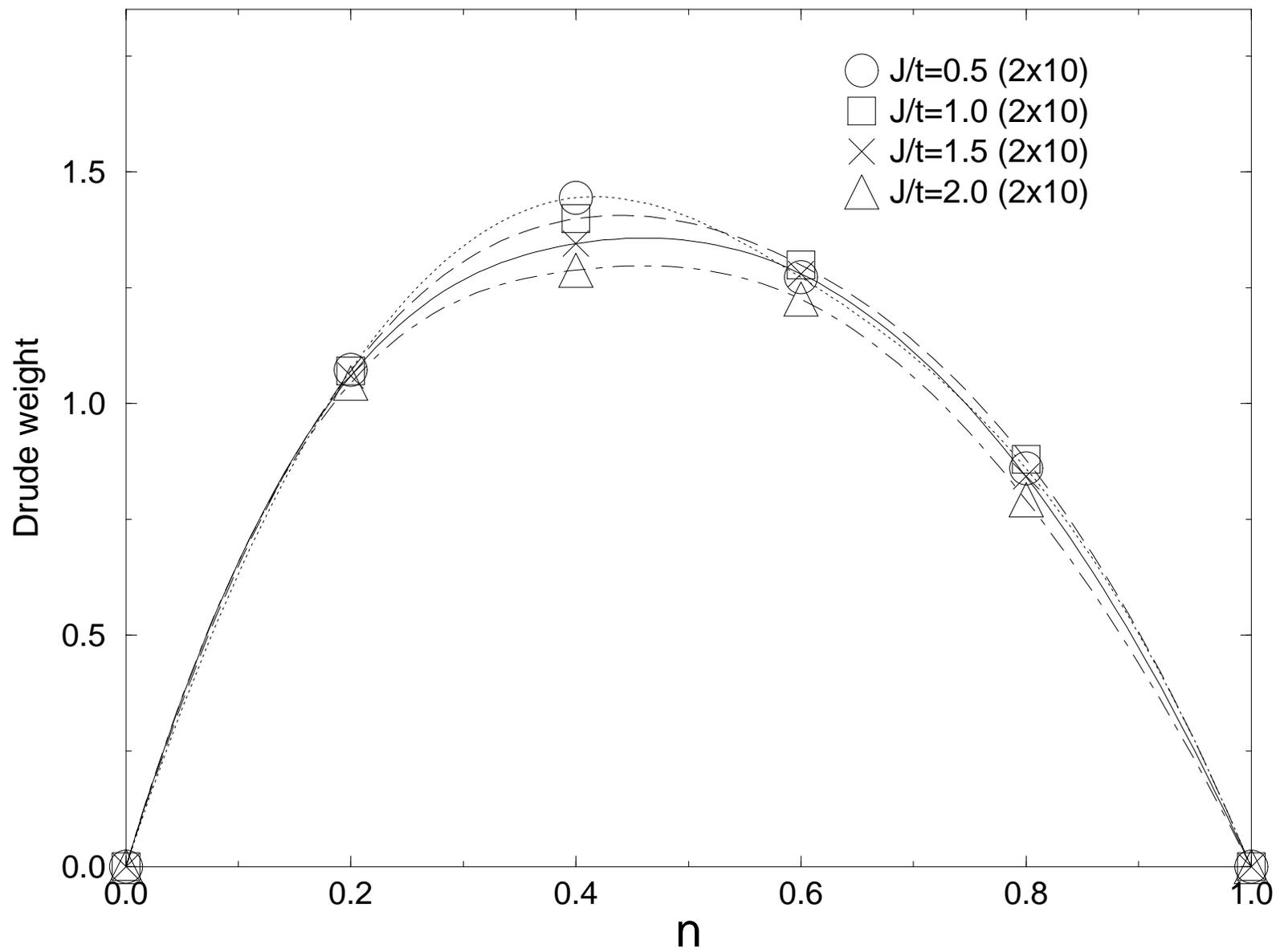
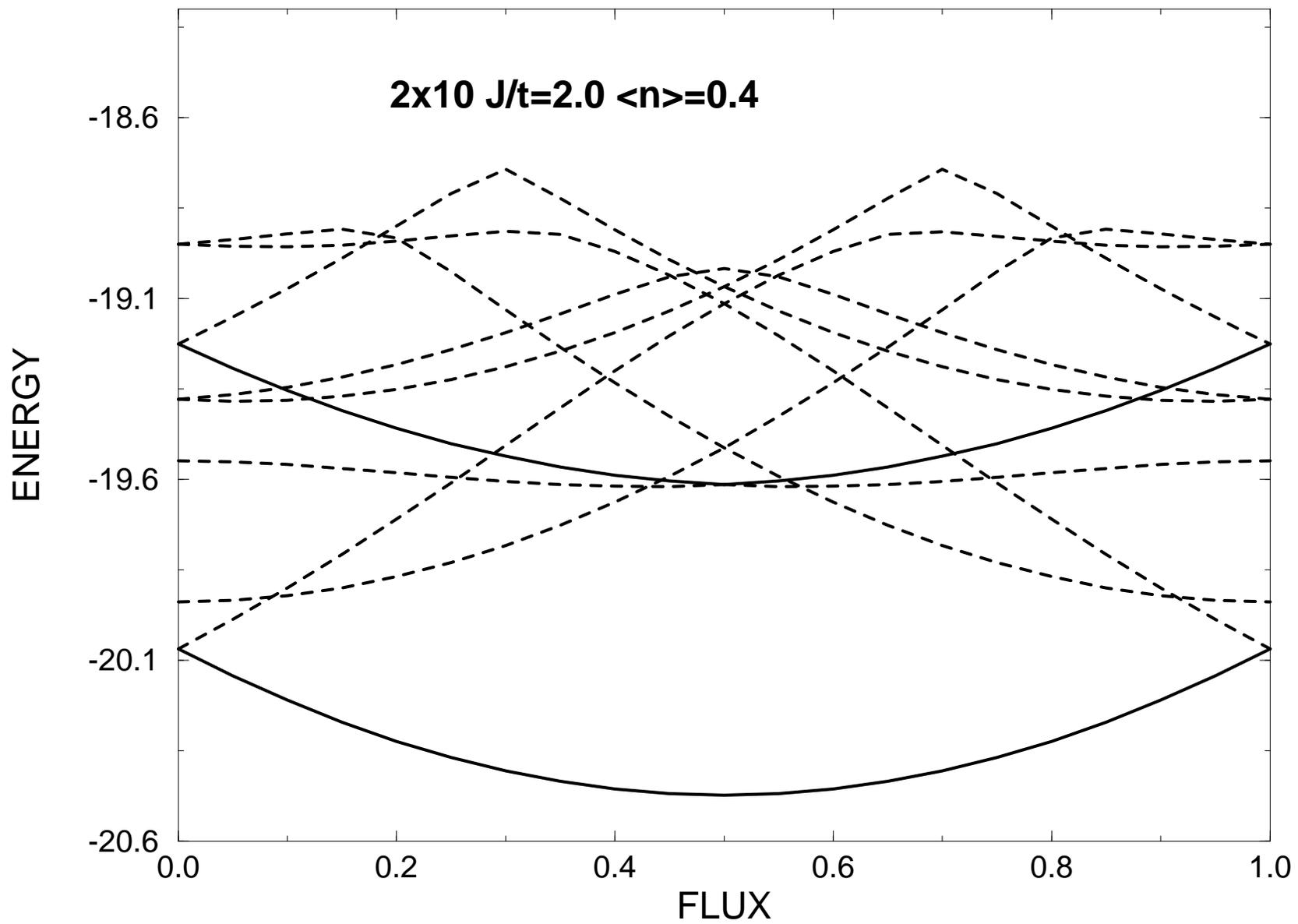
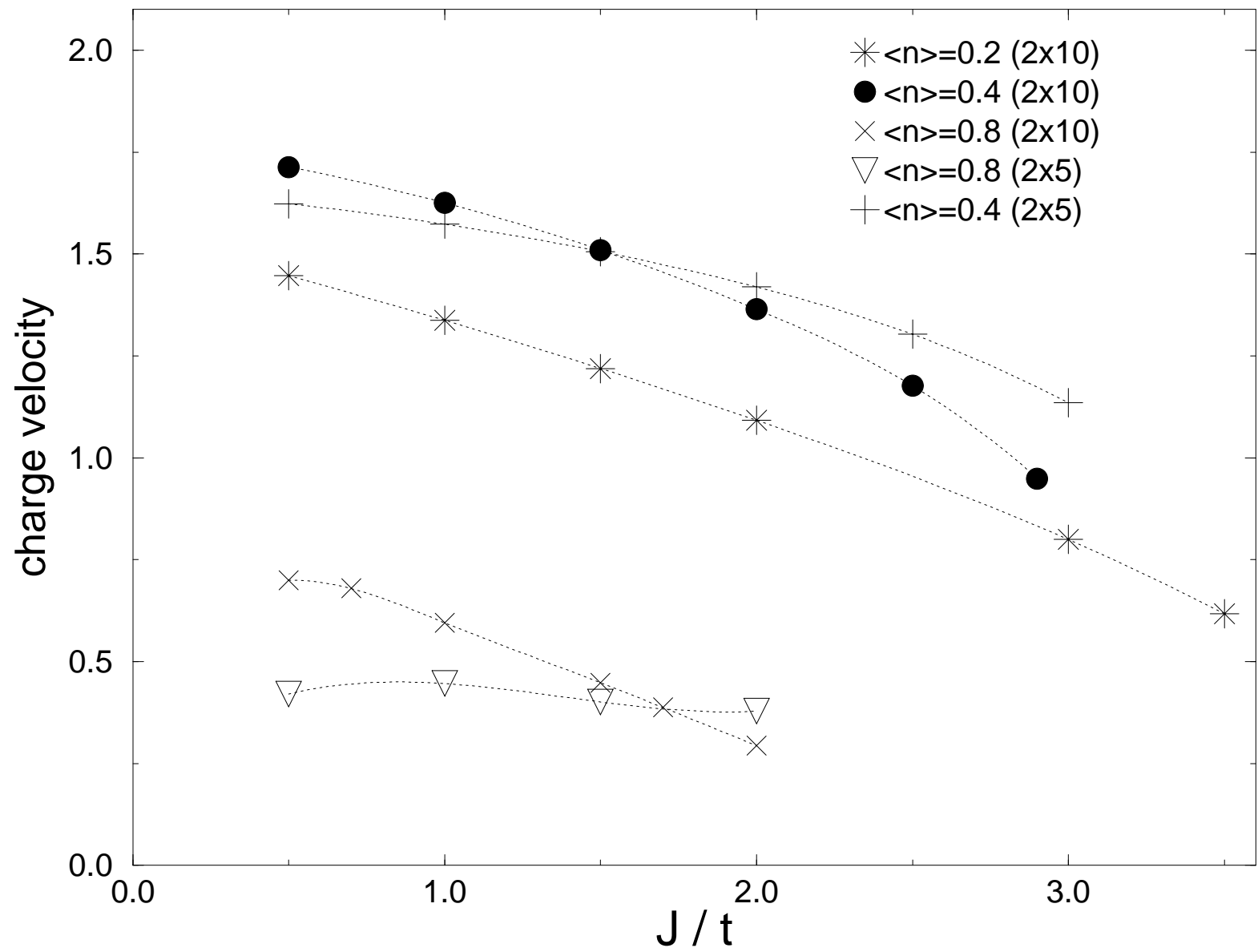
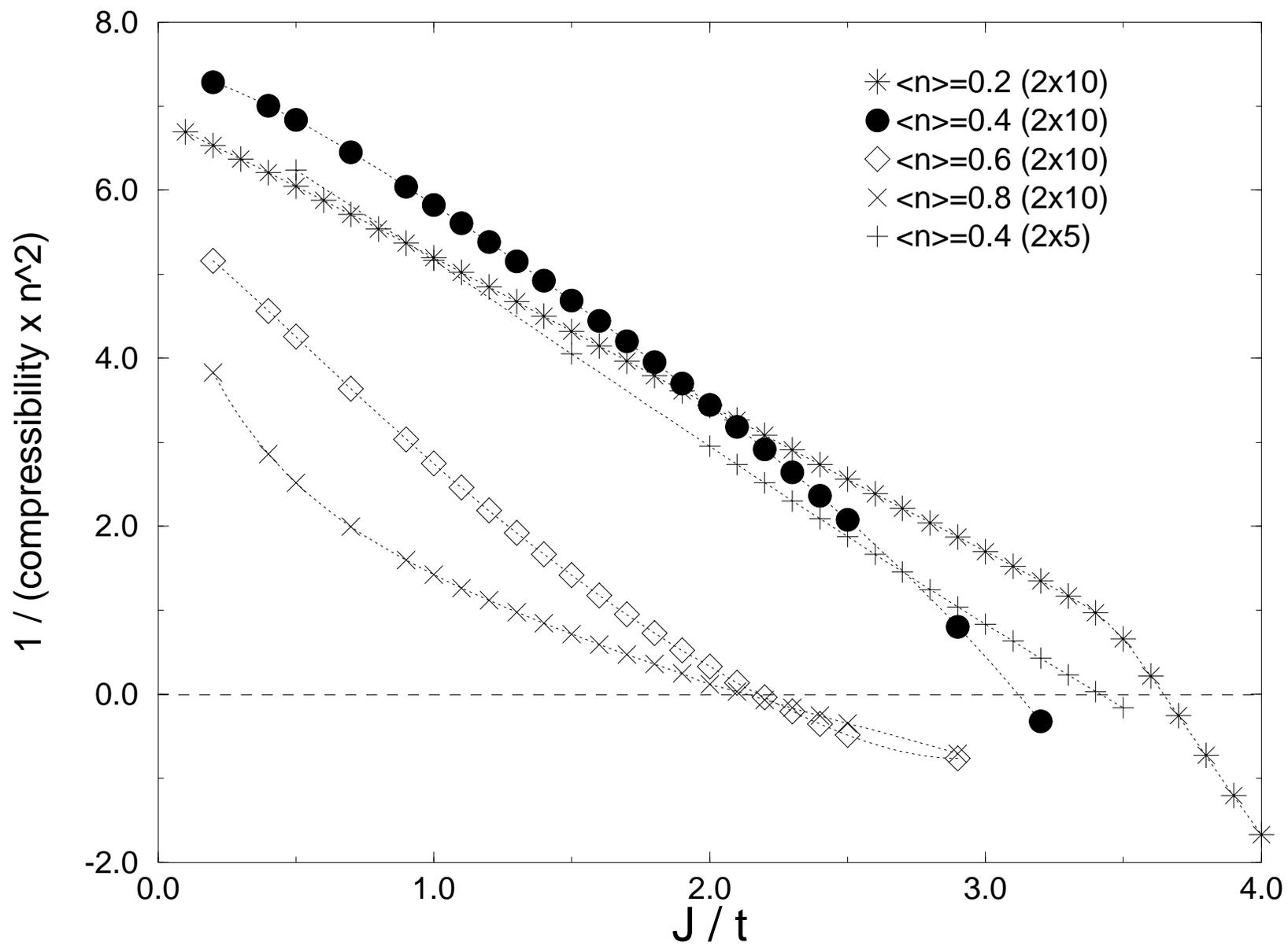
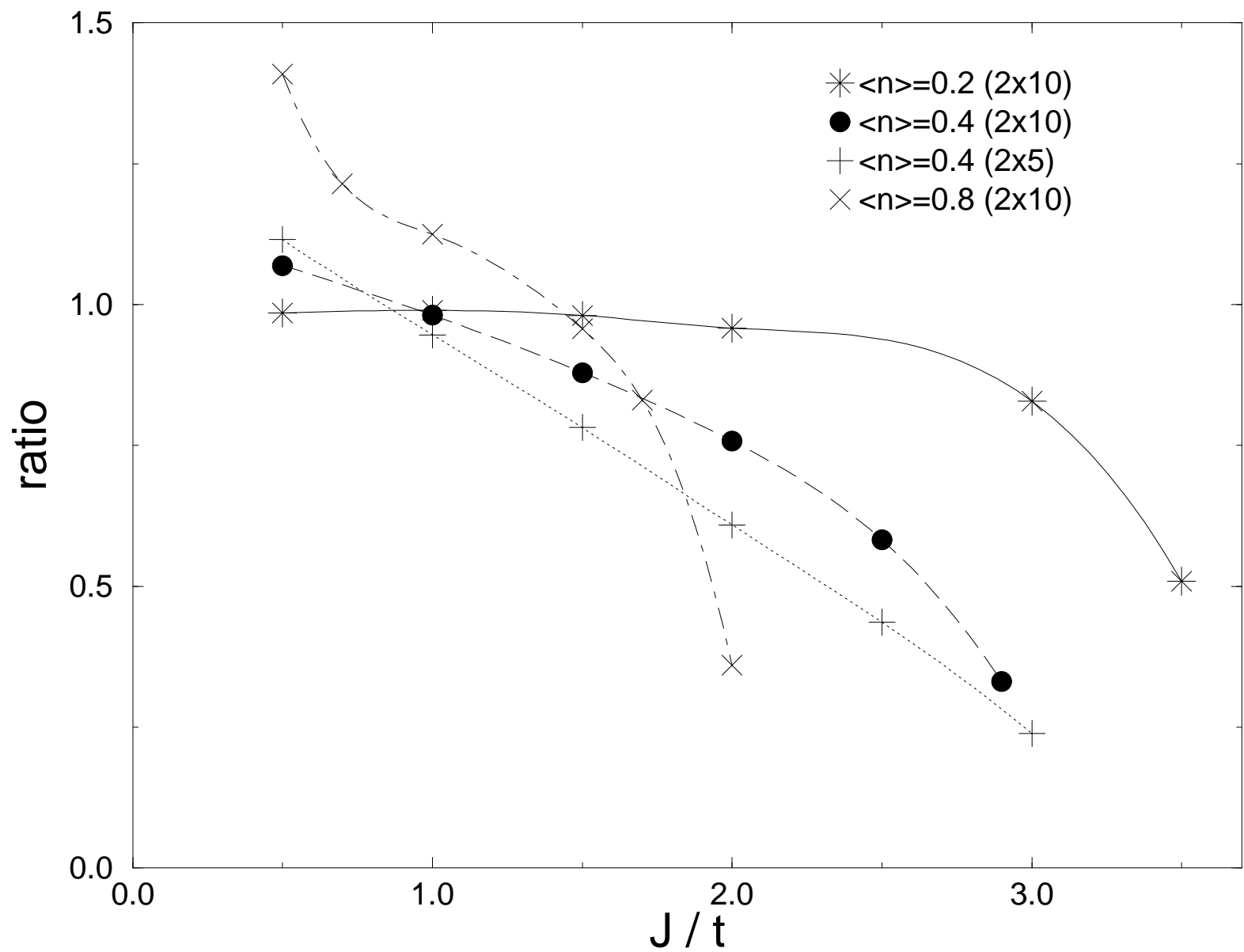
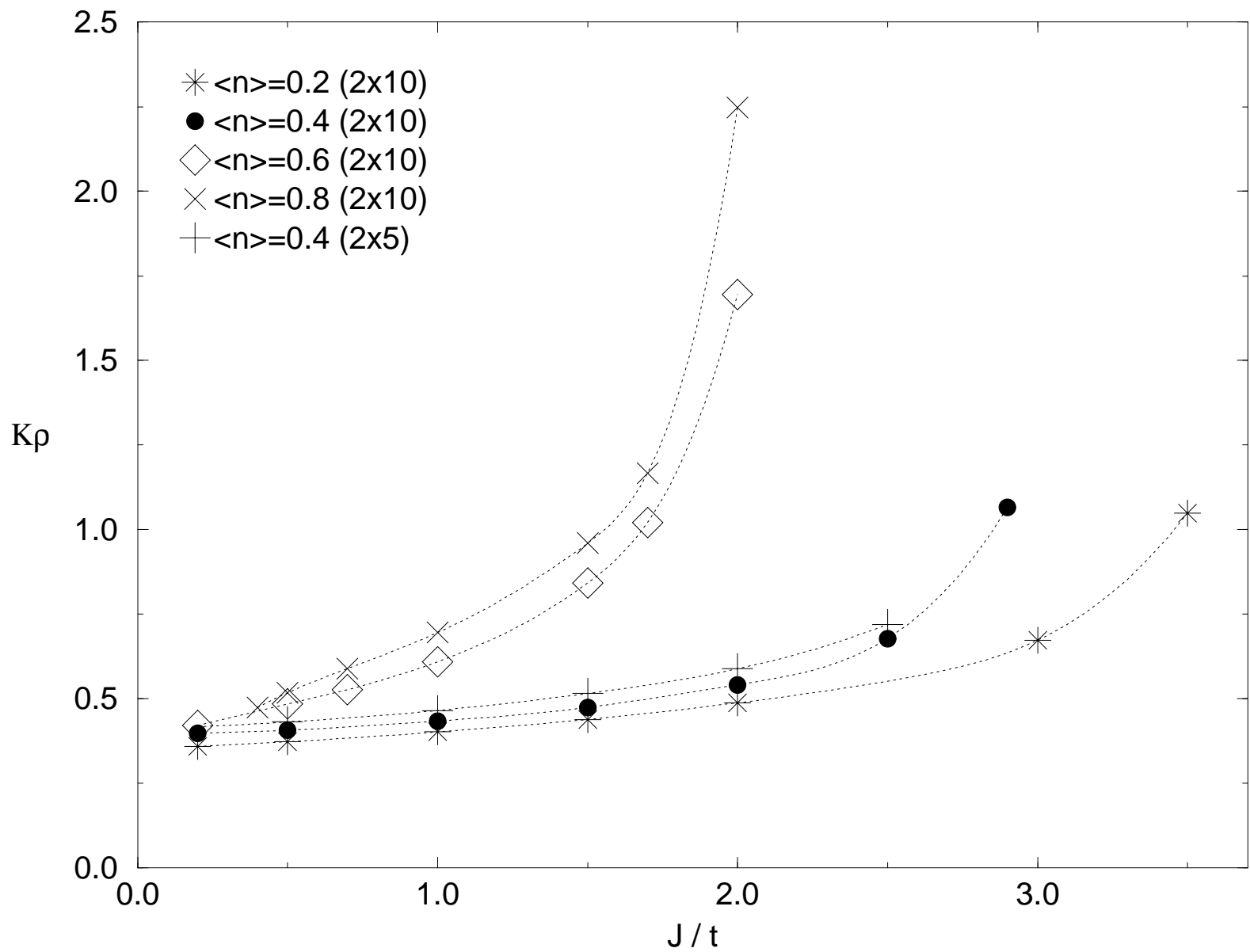
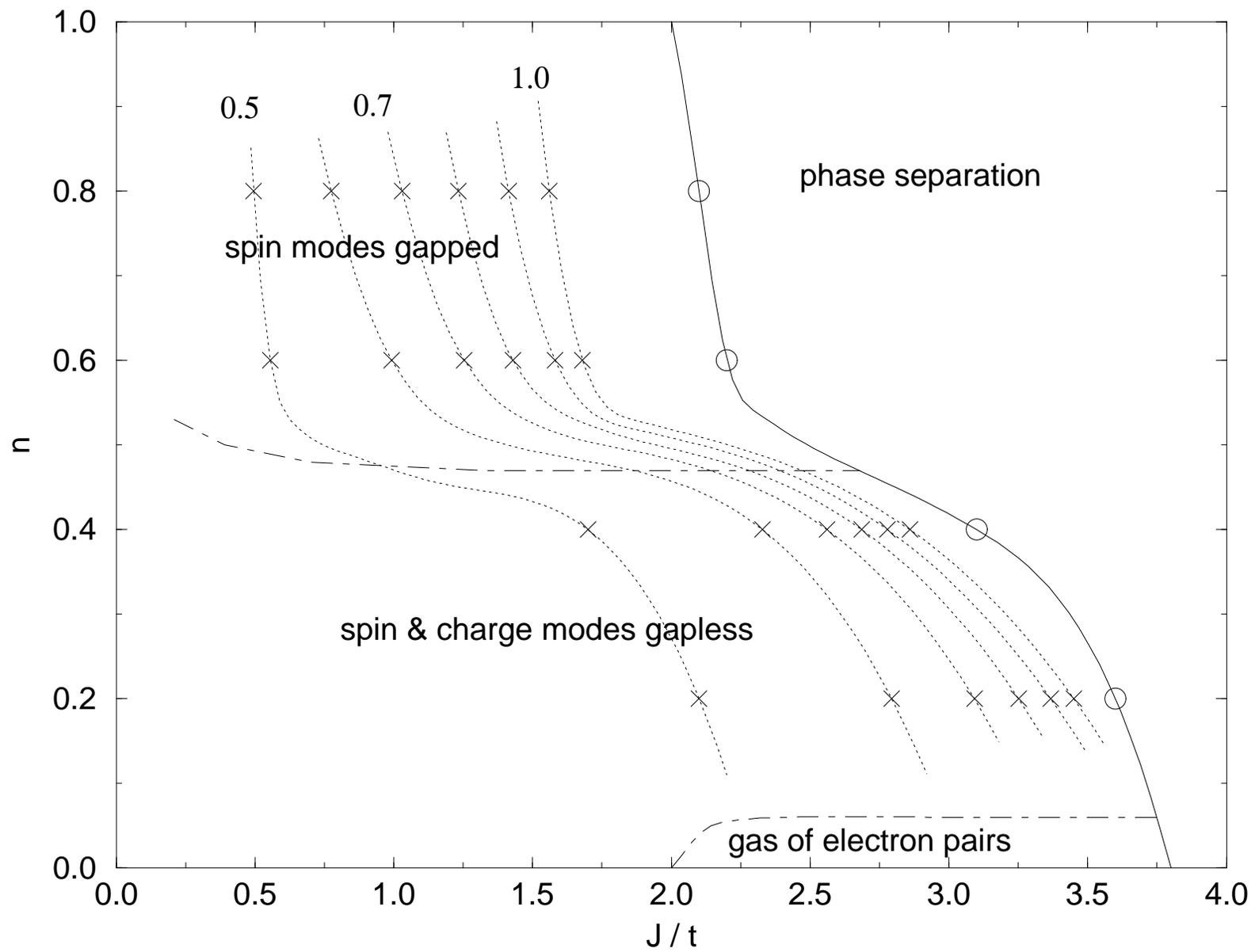
\end{document}